\documentclass[aps,prl,reprint,groupedaddress,amsmath,amssymb]{revtex4-1}
\usepackage{amsmath}
\usepackage{graphicx}
\usepackage{epstopdf}
\usepackage{color}
\usepackage[sort&compress]{natbib}

\begin{document}
	
	\title{Point-like inclusion interactions in tubular membranes}
	\author{Afshin Vahid$^1$, Timon Idema$^1$}
	\email[]{T.Idema@TUDelft.nl}
	\affiliation{\small \em $^1$Department of Bionanoscience, Kavli Institute of Nanoscience, Delft University of Technology, Delft, The Netherlands}

	\date{\today}
	
	\begin{abstract}
		We analytically study membrane mediated interactions between inclusions embedded in a tubular membrane. We model inclusions as constraints coupled to the curvature tensor of the membrane tube. First, as special test cases, we analyze the interaction between ring and rod shaped inclusions. Using Monte Carlo simulations, we further show how point-like inclusions interact to form linear aggregates. Our results reveal that depending on the hard-core radius of the inclusions, they arrange into either lines or rings to globally minimize the curvature energy of the membrane.	
	\end{abstract}
	
	
	\maketitle

	\textit{Introduction}.\textbf{\textemdash}Membrane nanotubes can be extracted experimentally from \textquoteleft giant\textquoteright \space unilamellar vesicles (GUVs) by different techniques like optical tweezers \cite{koster2005force} or micropipettes \cite{evans1996biomembrane,karlsson2001molecular,powers2002fluid}. In vivo, for example in the endoplasmic reticulum, these membrane tubes are generated either by being pulled out by molecular motors \cite{roux2002minimal} or pushed out by polymerizing cytoskeletal filaments \cite{shibata2009mechanisms}. The formation mechanism and the stability of tubular membranes have been extensively studied both theoretically \cite{bar1994instability,derenyi2002formation,fournier2007critical,monnier2010long} and experimentally \cite{evans1996biomembrane,karlsson2001molecular,koster2005force,polishchuk2003mechanism}.
	
	In addition to direct interactions like electrostatic forces, inclusions (like proteins) embedded in biological membranes experience interactions mediated by the elastic deformation of that membrane. Inclusions create these deformations by imposing a curvature field in the lipid bilayer when they are bound to or embedded in a membrane. Despite the presence of a repulsive pair potential between such inclusions in a flat membrane \cite{goulian1993long,weikl1998interaction}, because of the non-pairwise additive nature of many-body interactions, they collectively attract each other and form stable spatial patterns \cite{kim1999many}. Numerous analytical investigations \cite{fournier2003dynamin,auth2009budding} and computer simulations \cite{atilgan2007shape,reynwar2007aggregation} have been done to show that this non-additivity drives vesiculation and budding in biological membranes. In contrast to flat membranes, membrane-mediated interactions between inclusions embedded in tubular membranes are not well understood. These interactions can be found, for example, in the last step of exocytosis and in cell division, where some specific proteins make energy-favorable structures to facilitate  membrane scission \cite{morlot2013mechanics}. Compared to the scale of the plasma membrane which can be approximately considered as a flat surface, the curved nature of such a tubular membrane can significantly affect these interactions.
	Recently, it has been revealed that hard particles and semi flexible polymers absorbed to soft elastic shells, collectively induce aggregates and produce a rich variety of aggregation patterns \cite{reynwar2007aggregation,idema2010membrane,reynwar2011membrane,pamies2011reshaping,vsaric2011particle,zhang2012ordered,vsaric2013self,ramakrishnan2014mesoscale}. Particularly, P{\`a}mies and Cacciuto showed that spherical nanoparticles adhering to the outer surface of an elastic nanotube can self-assemble into diverse aggregates \cite{pamies2011reshaping}.
	They considered elastic nanotubes as stretchable and bendable structures; in contrast biological membranes cannot withstand shearing forces and are stretch free interfaces. Therefore, an obvious question to ask is what kinds of structure inclusions might induce in a cylindrical fluid surface.
	
	The aim of this paper is to analytically study the interactions between inclusions embedded in a membrane tube. We treat inclusions as point-like constraints imposing local curvature on the membrane. Previous work done by Dommersnes and Fournier \cite{dommersnes1999n,dommersnes2002many} already suggested a methodology to derive inclusion interactions mediated by membrane deformations in planar geometries. Using this framework, one can easily calculate the interaction of many point-like inclusions in a non-additive way. Here, we apply that framework to a membrane tube containing an arbitrary number of inclusions. For simplicity we assume that inclusions do not undergo any conformational changes, though these could also be accounted for using the same formalism~\cite{fournier2014dynamics}. After giving a brief outline of the model, first we look at some specific shapes like rings and rods, and afterwards we will study interactions between point-like inclusions. Using Monte Carlo simulations, we investigate the effects of different parameters like the density and the size of inclusions on their final equilibrium configuration. 
	
	Our results reveal that in contrast to the interaction of two rings, two infinite rods embedded in a membrane tube behave completely different from the same inclusions in a flat membrane. While two identical inclusions always repel each other in a flat membrane, in a cylindrical membrane they can also attract. We find a similar behavior for identical point-like particles, which can also attract and repel on the tube, depending on their separation and relative orientation. Consequently, for many inclusions, and depending on their hard-core radius, they form either ring or line like structures. We conclude that rings of membrane inclusions, such as the dynamin rings found in endocytosis, or the FtsZ rings found in bacterial cytokinesis~\cite{shlomovitz2009membrane}, can thus spontaneously form on tubular membranes, due to membrane-mediated interactions alone.
	
	\textit{Model}.\textbf{\textemdash}As mentioned earlier, we use the theoretical framework introduced in ref. \cite{dommersnes2002many}. We apply this method to membranes with a cylindrical topology. The unperturbed system is a perfect cylinder, parametrized by angular ($\theta$) and longitudinal ($\zeta = Z/R$, with $R$ the radius of the cylinder) coordinates. We describe deviations from the perfect cylindrical shape using the Monge gauge:
	\begin{equation}\label{Surface}
	\mathbf{r}(\theta,\zeta)= R \begin{pmatrix}
	(1+ u(\theta,\zeta))\cos(\theta)\\
	(1+ u(\theta,\zeta))\sin(\theta)\\
	\zeta
	\end{pmatrix},
	\end{equation}
	where $u(\theta,\zeta)<<1$. In order to mathematically describe the biological membrane, we use the Canham-Helfrich \cite{canham1970earlier, helfrich1973elastic} energy functional
	\begin{align}\label{eq:ShapeE}
	E = \int_{S} \mathrm{d}A \left(2\kappa H^{2} +\sigma \right),
	\end{align} 	
	where $\mathrm{d}A$, $\kappa$, $H$ and $\sigma$ are the surface element, bending rigidity, mean curvature, and surface tension, respectively. It is well known that, under the application of a constant force $f = 2\pi \sqrt{2 \kappa \sigma}$ to the membrane, a cylindrical tube of radius $R = \sqrt{\kappa / 2\sigma}$ is an equilibrium shape minimizing the energy functional given by Eq.~\ref{eq:ShapeE}~\cite{evans1996biomembrane,derenyi2002formation}.
	
	Following the construction by Dommersnes and Fournier, we put $N$ inclusions in the membrane at positions  $\left( \mathbf{r}_1,\mathbf{r}_2,...,\mathbf{r}_p,...,\mathbf{r}_N\right )$ imposing the curvature matrix $\mathbf{C} = \left(...,C_{\theta\theta}^p,C_{\zeta\theta}^p,C_{\zeta\zeta}^p,... \right)$, where $C_{ij}^p = \partial_{ij} u({\theta, \zeta) } \delta(\theta-\theta_p,\zeta-\zeta_p)$. To get the deformation field of the tube, $u(\theta,\zeta)$, we minimize the energy functional (Eq. \ref{eq:ShapeE}) given that we have imposed the curvature constraints. For the details of solving the resulting Euler-Lagrange equations please see the Supplemental Material~\footnote{See the \textit{Model} section in the Supplemental Material for the derivation}.
	In the case of self-interactions, we need to take the actual size of the inclusions into account, and should therefore introduce two cutoff wave vectors (we cannot have fluctuations with wavelength smaller than the size of the lipids): $\Lambda_\zeta = 1/a$ and $\Lambda_\theta = 2 \pi R / a $, where the cutoff radius $a$  is chosen such that $\Lambda_{\theta (\zeta)}^{-1}$ is in the order of the membrane thickness \cite{barbetta2010surface}. 
	
	Using this formalism, we can get an analytical expression for the elastic energy and the shape of the deformed membrane for any arbitrary number of inclusions. The nondimensionalized components of the curvature tensor $\mathbf{C}$, for a tube with a thickness of $\simeq 5$~nm and radius $\simeq 20-50$~nm, are in the order of $c^{-1}\simeq 0.1-0.25$. In the following, we measure the energy in units of $2 \pi \kappa c^2$, which, for the standard values of $ \kappa= 30 k_\mathrm{B} T$  and $c=10$, equals
	$2 \pi \kappa c^2 \simeq 20\times 10^3 k_\mathrm{B} T $.\\
	
\begin{figure}[t]
		\includegraphics[width=0.49\textwidth]{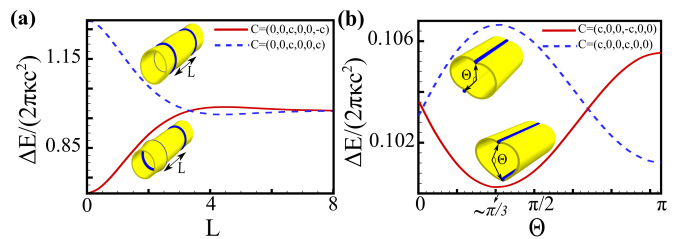}
		\caption{The calculated energy cost of having two inclusions (as compared to none) for a membrane tube as a function of the distance between (a) two rings and (b) two rods. Inclusions impose either the same (dashed line) or opposite (solid line) curvatures.}
		\label{fig1_RingsAndRods} 
	\end{figure}

	\textit{Special test cases}.\textbf{\textemdash} To show the difference between planar and highly curved regimes, we study two special shapes of inclusions using the described formalism. First, we look at the interaction between two rings, separated by a distance $L$, in a cylindrical membrane (Fig.~\ref{fig1_RingsAndRods}a). Second, we analyze the energy favorable configuration of rod-like inclusions embedded in a membrane tube (Fig.~\ref{fig1_RingsAndRods}b). By considering ring shaped constraints, recent studies have constructed a variational framework to model the constriction process during cytokinesis \cite{almendro2013mechanics,almendro2015analytical}. Also, using an analytical approach, the wrapping process of a rod like particle by a tubular membrane has been studied via minimization of bending and adhesion energies \cite{chen2010adhesion}.
	
	The energy dependence on inclusion separation between two rings is shown in Fig. \ref{fig1_RingsAndRods}a. We find that two identical rings ($\mathbf{C}=(0,0,c,0,0,c)$) have strong short range repulsion and weak long range attraction; this behavior causes two rings imposing equal curvature to not coalesce, but equilibrate at a certain distance from each other. The long-range attraction originates from the fact that the membrane's size is finite in the angular direction, resulting in a reduction in the total energy of two overlapping tails when distant rings move closer together. For different radii of the tube, we get different equilibrium separations for the rings; the larger the radius is, the further the rings are away from each other (Fig.~S1~\footnote{See Fig. S1 in the Supplemental Material}). The situation for rings imposing opposite curvature will be reversed. The membrane, to globally minimize its curvature energy, favors two rings to coalesce despite having a local minimum for larger separations.	
	
	In contrast to rings, two rods interact completely differently. Depending on their angular separation ($\Theta$), two identical rods ($\mathbf{C}=(c,0,0,c,0,0)$) can either attract or repel each other (Fig. \ref{fig1_RingsAndRods}b). One clear difference with both flat membranes and the previous test case is that the tails of deformations in the angular direction are limited to a confined space and overlap. Consequently, there are two contributions to the total energy of the tube: one is due to the membrane deformation between two rods and the other one originates from the overlapping tails. For small distances, these two interactions add to a net attraction between identical rods, as this minimizes the overlap between their tails. For larger separations, the effect of the deformed membrane between the inclusions becomes dominant, and in order to minimize the bending energy of the system, they sit on the opposite poles. Similar to rings, the location and the strength of the energy barrier depends on the radius of the tube. In the limit of very large $R$, the interaction between two rods imposing the same curvature is purely repulsive (Fig.~S2~\footnote{See Fig.~S2 in the Supplemental Material}), like in a flat membrane~\cite{muller2007balancing}.
	\begin{figure}[t]
		\includegraphics[width=0.32\textwidth]{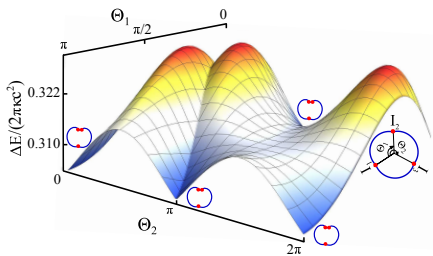}
		\caption{The energy landscape for a membrane tube containing three rod like inclusions I1, I2 and I3.}
		\label{fig2_ThreeInclusions}
	\end{figure}
	Since membrane mediated interactions, in contrast to for example electrostatic interactions, behave in a non-additive way, it is interesting to look at a system with more than two inclusions. Particularly, we find that adding a third rod into the previous system makes the repulsion between the first two attractive. The global minimum of the three dimensional energy landscape, as illustrated in Fig.~\ref{fig2_ThreeInclusions}, corresponds to the situation that two rods are on top of each other and the third one is on the opposite pole. Similarly, for more than three inclusions, we find that for an even number of rods the global minimum occurs when they equally distribute between the two poles; and in case of having an odd number of inclusions, one of the poles will have one more rod than the other.

\begin{figure}[t]
		\includegraphics[width=0.41\textwidth]{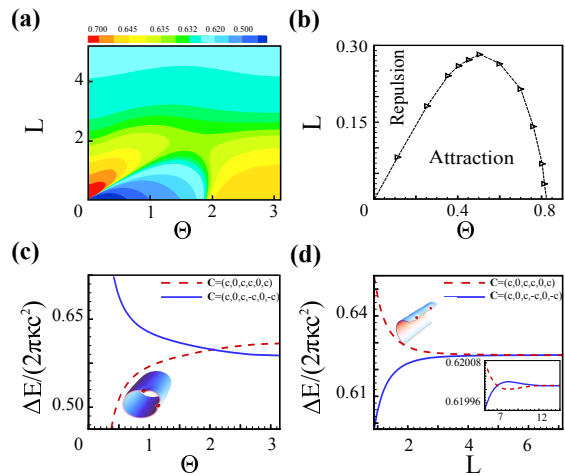}		
		\caption{(a) The curvature energy ($\frac{\Delta E}{2\pi \kappa c^2}$) of a membrane containing two inclusions, as a function of their angular ($\Theta$) and longitudinal ($L$) separation, with $L$ in units of the tube radius $R$. (b) The line around the global minimum at which the energy equals the local minimum at large separations. For particles whose diameter exceeds the size of this region, the overall behavior is repulsive (settling in the local minimum at large separations). Smaller particles globally attract, but have a high energy barrier separating the attractive and repulsive regime. (c) Two identical inclusions placed at the same longitudinal coordinates $(L=0)$ attract each other. (d) Point-like inclusions behave like rings when they are situated on the same transversal coordinates ($\Theta = 0$); the inset shows the weak long-rage attraction.}
		\label{fig3_TwoPointLikeInclusions}
	\end{figure}

	\textit{Point-like inclusions}.\textbf{\textemdash} Before focusing on many body interactions, let us  first consider a tubular membrane containing two identical inclusions imposing the same curvature, so $\mathbf{C}=(c,0,c,c,0,c)$ (similar to rods and rings, the behavior for inclusions inducing opposite curvature will be reversed). Fig.~\ref{fig3_TwoPointLikeInclusions}a depicts the excess curvature energy of the membrane as a function of both angular and longitudinal distances between two inclusions. 
	At small distances there are two different kinds of behavior corresponding to two distinct directions: along the tube axis two inclusions strongly repel each other at short distances and attract each other at longer distances (Fig.~\ref{fig3_TwoPointLikeInclusions}d), while in the transversal direction the two-body interaction is purely attractive (Fig.~\ref{fig3_TwoPointLikeInclusions}c). When two identical point-like inclusions have the same transversal coordinates ($\Theta = 0$), they behave like rings, although the long-range attraction becomes very weak (see inset in Fig.~\ref{fig3_TwoPointLikeInclusions}d). However, when these inclusions have the same longitudinal coordinates ($L = 0$), their behavior differs from that of the infinite rods. While for the rods we find both short-range attraction and long-range repulsion, identical point-like inclusions at the same longitudinal coordinate always attract. The global energy minimum of the system corresponds to the two inclusions sitting next to each other in the angular direction (see Fig.~\ref{fig3_TwoPointLikeInclusions}a). However, if the inclusions are initially separated, there is a large energy barrier (on the order of $\sim100 k_\mathrm{B} T$) that the inclusions have to overcome to reach this global minimum state. Moreover, the region around the global minimum where the energy is less than that at the local minimum at large separations (see inset in Fig.~\ref{fig3_TwoPointLikeInclusions}d) is only very small, as shown in Fig.~\ref{fig3_TwoPointLikeInclusions}b. Consequently, small inclusions globally attract, but may not find each other due to the large barrier; particles with a diameter larger than the size of the attractive basin in Fig.~\ref{fig3_TwoPointLikeInclusions}b have a global minimum at large but finite separation, also separated from the (now local) minimum close together by a large barrier.
	
	Like for rods, adding more inclusions changes the energy landscape. For point-like inclusions the net effect is a lowering of the barrier between the energy minima at small and large separations. Consequently, the presence of other inclusions can allow two inclusions to reach their global equilibrium state, which could potentially take very long if those other inclusions were absent. 
	
\begin{figure}[t]
		\includegraphics[width=0.48\textwidth]{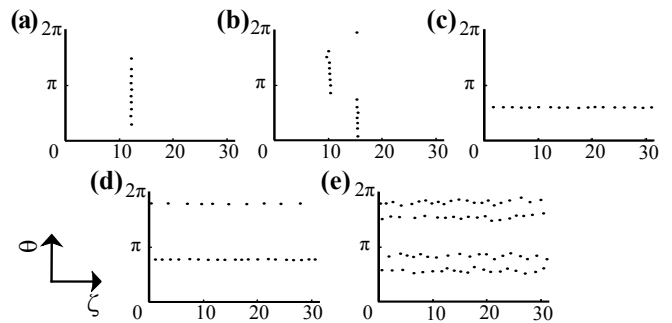}
		\caption{Equilibrium configurations obtained by Monte Carlo simulation for a system containing  (a) 10 inclusions with hard-core radius of $a_0= 0.2$ (b) 16 inclusions with hard-core radius of $a_0= 0.2$ (c) 16 inclusions with hard-core radius of $a_0= 1.1$ (d) 30 inclusions with hard-core radius of $a_0= 1.1$ (e) 80 inclusions with hard-core radius of $a_0= 1.1$. \label{fig4_ManyBodyInteractions}}
	\end{figure}

	To elucidate the collective behavior of multiple inclusions packed in the system, we perform Monte Carlo (MC) simulations on a membrane tube containing inclusions with different hard-core radii (which are introduced to take into account the finite size of the particles). During the simulations, we consider periodic boundary conditions in the longitudinal direction. The only effect of a non-zero hard-core radius of inclusions is the transition from the short-range attractive-dominated regime to the repulsion dominated area. In all cases the tube's reduced length is $\zeta = 10 \pi$ and correspondingly, the cut-off wave vectors  are $\Lambda_{\zeta}=314$ and $\Lambda_{\theta}=62$ for the cutoff radius of $a= 0.1$.  During MC simulations we use the Metropolis algorithm \cite{metropolis1953equation} with parallel tempering \cite{hansmann1997parallel}. As membrane mediated interactions between inclusions originate from both the average deformation of the membrane and the constraints imposed on its shape fluctuation, one may be concerned about the Casimir interactions. In our system, the thermal fluctuation effects nicely decouple from the elastic ones \cite{li1992fluctuation}, and it is straightforward to show that their effects are relatively small, quickly fading out with the distance between inclusions \cite{lin2011fluctuation} (Fig.~S3~\footnote{See Fig. S3 in the Supplemental Material}).
	We find that for an arbitrary number of inclusions with a hardcore radius $a_0= 0.2$, they will attract each other in the angular direction and self-assemble into ring like configurations (Figs. \ref{fig4_ManyBodyInteractions}a and \ref{fig4_ManyBodyInteractions}b). Because of having a rough energy landscape, including many barriers like the one shown in Fig.~\ref{fig3_TwoPointLikeInclusions}a, inclusions could not always completely merge and reach the global energy minimum. However, we can certainly conclude that in order to minimize the curvature energy of the membrane, such identical inclusions will assemble into rings.
	This process is reminiscent of recruiting dynamin proteins during exocytosis, during which they self-assemble and form rings to constrict the membrane and, finally, separate the nascent vesicle  from the cell. In contrast, for inclusions having a larger radius ($a_0=1.1$), our MC simulations reveal that they collectively align in the longitudinal direction. Therefore, as shown in Fig. \ref{fig4_ManyBodyInteractions}c, if the number of particles is less than that fits the length of the tube they aggregate into one line. The boundary for which the transition from rings to lines occurs is shown in Fig. \ref{fig3_TwoPointLikeInclusions}b: if the radius of inclusions is such that it cannot fall in the attractive area, they self-assemble into lines. If we increase the particle density (Figs. \ref{fig4_ManyBodyInteractions}d and \ref{fig4_ManyBodyInteractions}e), such that they do not all fit on a single line anymore, they do not make other configurations, but distribute around two lines on the opposite poles. The reason for this is actually hidden in the assumptions of the theoretical model we use. First, inclusions are treated as point like constraints that impose a uniform curvature in all directions. Second, while as in our model, a fluid membrane cannot resist any stretch, it has recently been shown that in an elastic membrane the competition between bending and stretching rigidities gives rise to different configurations like helical structures \cite{pamies2011reshaping}; in the limit of very small stretching rigidity, linear aggregations like rings and rods are the only configurations that one can get for an elastic tube.
	
	\textit{Conclusion}.\textbf{\textemdash}We have investigated the curvature mediated interactions between different identical  inclusions. We have shown that while rings have strong short-range repulsion (and weak long-range attraction), identical rods can either attract or repel each other depending on the angular distances between them. For two point like inclusions embedded in a tubular membrane, our analytical solutions show that they attract and repel each other in the transversal and longitudinal direction, respectively. 
	Our study of a  membrane tube containing many inclusions has highlighted the importance of many body interactions for the inclusions in order to collectively induce aggregations. Having done Monte Carlo simulations on such a system,  we observed that depending on the defined hard core radius, inclusions self-assemble into line or ring like structures. The results may explain the mechanisms by which inclusions self-assemble during membrane constriction in the processes like exocytosis and cytokinesis. 
	
	\begin{acknowledgments}
		We would like to thank Dr. J. L. A. Dubbeldam for fruitful discussions. This work was supported by the Netherlands Organisation for Scientific Research (NWO/OCW), as part of the Frontiers of Nanoscience program.
	\end{acknowledgments}

\end{document}


\title{Supplementary Material for\\ ``Point-like inclusion interactions in tubular membranes''}

	\author{Afshin Vahid$^1$, Timon Idema$^1$}
	\affiliation{\small \em $^1$Department of Bionanoscience, Kavli Institute of Nanoscience, Delft University of Technology, Delft, The Netherlands}
	
	\date{\today}
	
	\maketitle	
	
	\section{Model}
	We use the methodology developed by Dommersnes and Fournier [27, 28]. As stated in the main text, we apply this method to membranes with a cylindrical topology. The unperturbed system is a perfect cylinder, parametrized by angular ($\theta$) and longitudinal ($\zeta = Z/R$, with $R$ the radius of the cylinder) coordinates. We describe deviations from the perfect cylindrical shape using the Monge gauge:
	\begin{equation}\label{Surface}
	\mathbf{r}(\theta,\zeta)= R \begin{pmatrix}
	(1+ u(\theta,\zeta))\cos(\theta)\\
	(1+ u(\theta,\zeta))\sin(\theta)\\
	\zeta
	\end{pmatrix},
	\end{equation}
	where $u(\theta,\zeta)<<1$ and $R = \sqrt{\kappa / 2\sigma}$. Assuming that $u(\theta,\zeta)$ is sufficiently differentiable, we calculate the mean curvature $H$ and surface element $\mathrm{d}A$ as 
	\begin{widetext}
		\begin{align}
		H &= 
		\frac{-2 u_{\zeta } u_{\theta } u_{\theta \zeta }-(1+ u_{\zeta }^2 )\left(-u_{\theta \theta }+u+1\right)+ u_{\zeta \zeta } \left(u_{\theta }^2+(u+1)^2\right)-\frac{2 u_{\theta }^2}{(u+1)}}{2 R \left((u+1)^2 \left(u_{\zeta }^2+1\right)+u_{\theta }^2\right){}^{3/2}}, \label{meancurvature}\\
		\mathrm{d}A &= R^2 (u+1) \sqrt{\left(u_{\zeta }^2+1\right)+u_{\theta }^2} \mathrm{d}\theta \mathrm{d}\zeta,
		\end{align}
	\end{widetext} 
	where $u_\zeta=\partial u / \partial\zeta$ etc. Assuming $N$ inclusions in the membrane at the positions  $\left( \mathbf{r}_1,\mathbf{r}_2,...,\mathbf{r}_N\right )$ imposing the curvature matrix $\mathbf{C} = \left(...,C_{\theta\theta}^p,C_{\zeta\theta}^p,C_{\zeta\zeta}^p,... \right)$, where $C_{ij}^p = \partial_{ij} u({\theta, \zeta) } \delta(\theta-\theta_p,\zeta-\zeta_p)$, $p=1,...,N,$ the curvature energy functional becomes:
	\begin{align}
	E = \int_{S} \mathrm{d}A \left(2\kappa H^{2} + \sigma - \Lambda_{\alpha} C_{\alpha}\right),
	\end{align}
	where the $\Lambda_{\alpha}$ are $3N$ Lagrange multipliers and $\alpha = 1,...,3N$. Since we use a Monge gauge parameterization in which we assume that $u(\theta,\zeta)$ is very small, the topology of our system is invariant. We therefore disregard the Gaussian curvature contribution, because according to the Gauss-Bonnet theorem the integral over a surface of fixed topology is constant. We also assume that the spontaneous curvature, which describes the asymmetry of the membrane, is zero. Substituting $H$ and $\mathrm{d}A$ into the energy functional and minimizing it up to first order in $u(\theta, \zeta)$, we obtain:
	\begin{align}
	\label{pde}
	\left(\nabla ^4 +2 \partial _{\theta \theta}+1 \right) u\left({\theta,\zeta}\right) &= \Lambda_\alpha D_\alpha\left({\theta,\zeta}\right),
	\end{align}
	where $\nabla ^4 =  \partial _{\theta\theta \theta\theta} + 2 \partial _{\zeta \zeta\theta\theta}+ \partial _{\zeta \zeta \zeta \zeta} $ is the biharmonic operator in cylindrical coordinates, and
	\[ \mathbf{D} = \left( \delta_{\theta\theta}^1,\delta_{\zeta\theta}^2,\delta_{\zeta\zeta}^3,...,\delta_{\theta\theta}^{3N-2},\delta_{\zeta\theta}^{3N-1},\delta_{\zeta\zeta}^{3N} \right), \]
	with $\delta_{ij}^{\alpha} = \partial_{ij} \delta(\theta-\theta_\alpha,\zeta-\zeta_\alpha)$. Because equation~(\ref{pde}) is linear, we can solve it using superposition once we know the Green's function, for which we obtain: 
	\begin{align}
	\label{Greensfunction}
	G\left({\theta,\zeta}\right) = \sum\limits_{n\neq \pm 1}\frac{  \left(\frac{e^{-\zeta \alpha _-(n)}}{\alpha _-(n)}-\frac{e^{-\zeta  \alpha _+(n) }}{\alpha _+(n)}\right)}{4 \pi \left(\alpha _+(n){}^2-\alpha _-(n){}^2\right)}\cos(n \theta),
	\end{align}
	where $ \alpha _\pm (n)\text{=}\sqrt{n^2\pm \sqrt{2 n^2-1}}$. The solution of equation~(\ref{pde}) is then given by
	\begin{align}
	u(\theta, \zeta) = \Lambda_\alpha G_\alpha(\theta, \zeta),
	\end{align}
	where the elements of the vector $\mathbf{G}$ are given by $\mathbf{G} = \left(...,C_{\theta\theta}^p,C_{\zeta\theta}^p,C_{\zeta\zeta}^p,... \right)$. To relate the Lagrange multipliers $\Lambda_\alpha$ to the actual constraints $C_\alpha$, we rewrite equation~(\ref{pde}) as
	\begin{align}
	\mathcal{L}_\alpha u = C_\alpha,
	\end{align}
	which gives us $\mathcal{L}_\alpha (\Lambda_\beta G_\beta) = \mathcal{L}_\alpha (G_\beta) \Lambda_\beta = C_\alpha$. Defining $M_{\alpha \beta} = \mathcal{L}_\alpha (G_\beta)$, we get:
	\begin{align}
	\Lambda_\alpha &= M_{\alpha \beta}^{-1} C_\beta, \\
	u(\theta, \zeta) &= M_{\alpha \beta}^{-1} G_\beta(\theta, \zeta) C_\beta, \\
	E &= \frac{\kappa}{2} M_{\alpha \beta}^{-1} C_\alpha C_\beta.
	\end{align}
	In the case of self-interactions, we calculate the derivatives of the Green's function in Fourier space,
	\begin{widetext}
		\begin{align}
		G_{klrs}\left({0,0}\right) = \frac{1}{2\pi^2}\sum\limits_{n\neq \pm 1}^{ \Lambda _ \theta}\int_{0}^{\Lambda_\zeta}\frac{\partial^4}{\partial k\partial l\partial r\partial s}\left(\frac{ e^{i(q \zeta+ n\theta)}}{(q^2+n^2)^2-2n^2+1}\right)\left.\right|_{\zeta=0,\theta=0} \mathrm{d}q,
		\end{align}
	\end{widetext}
	where the indices $k$, $l$, $r$ and $s$ are either $\theta$ or $\zeta$, and the cutoff wavevectors are related to the membrane thickness $a$ through $\Lambda_\zeta = 1/a$ and $\Lambda_\theta = 2 \pi R / a$, as given in the main text.
	
	\section{Special test cases}
\begin{figure}[ht]
		\includegraphics[width=0.48\textwidth]{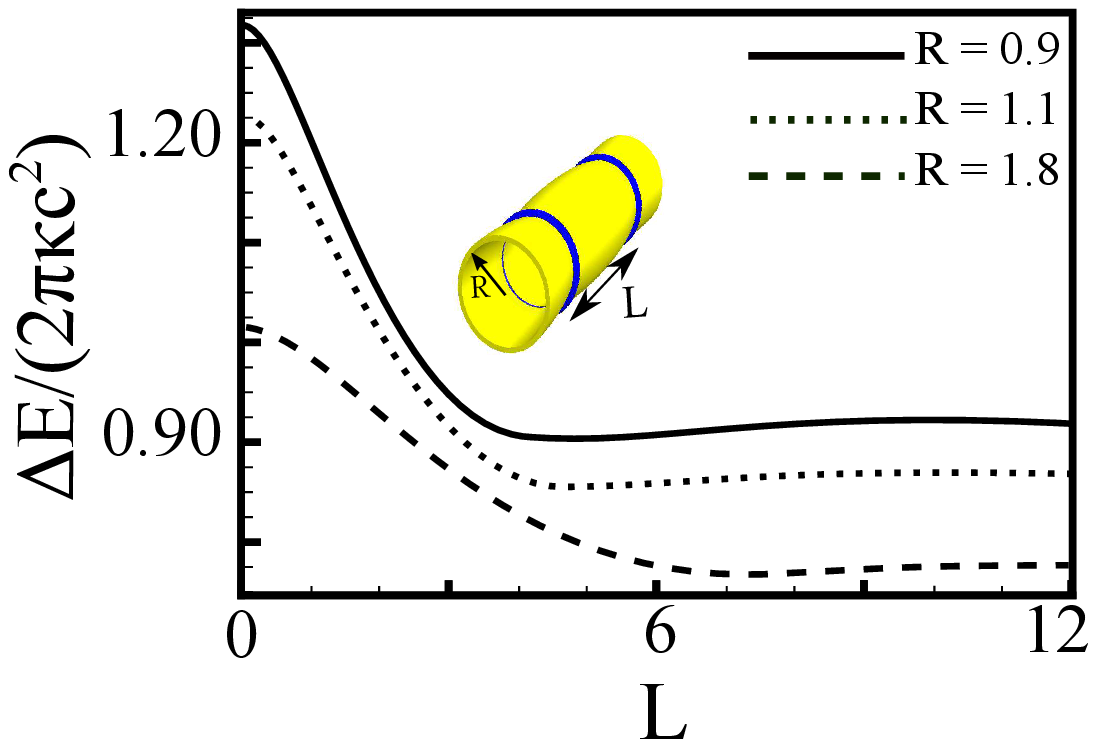}
		\caption{The competition between bending modulus and surface tension of the tube determines both the radius of the tube ($R = \sqrt{\kappa / 2 \sigma}$) and the equilibrium distance between two rings.}
		\label{figS1_RingsAndRods}		
	\end{figure}

	To evaluate the interaction between rings whose deformations depend only on the longitudinal coordinate ($\zeta$), we obtained simplified relations for one dimension. By letting the position vector of the membrane depend only on the longitudinal coordinate, we obtain the Green's function and the excess energy of the membrane between two rings:
	
	\begin{align}
	G(\zeta) \text{=}\frac{e^{-\left| \zeta \right| /\sqrt{2}}}{\sqrt{2}}\left[\sin \left(\frac{\left| \zeta \right| }{\sqrt{2}}\right)+\cos \left(\frac{\zeta }{\sqrt{2}}\right)\right],
	\end{align}
	\begin{widetext}
		\begin{align}
		& E(\Lambda_\zeta,L)= \frac{ \sqrt{2} \arctan \left( \frac{2\sqrt{2}\Lambda_\zeta}{(\Lambda_\zeta^2+1)^2-2} \right)-2\sqrt{2} \pi+4 \Lambda_\zeta  +4 \sqrt{2}\pi e^{-\frac{L}{\sqrt{2}}} \left(\sin \left(\frac{L}{\sqrt{2}}\right)+\cos \left(\frac{L}{\sqrt{2}}\right)\right)}{\frac{1}{16 \pi ^2} \left(\sqrt{2} \arctan \left( \frac{2\sqrt{2}\Lambda_\zeta}{(\Lambda_\zeta^2+1)^2-2} \right)-2\sqrt{2} \pi+4 \Lambda_\zeta \right)^2-2 e^{-\sqrt{2} L} \left(\sin \left(\sqrt{2} L\right)+1\right)},
		\end{align}
	\end{widetext}
	where $\Lambda_\zeta$ is the cutoff wave vector in the longitudinal direction. All the lengths are non-dimensionalized by expressing them in terms of the unperturbed tube radius~$R$; by plugging back $R$ into the equations we find that the equilibrium distance between the rings increases with $R$ (Fig.~\ref{figS1_RingsAndRods}).
	
\begin{figure}[ht]
		\includegraphics[width=0.48\textwidth]{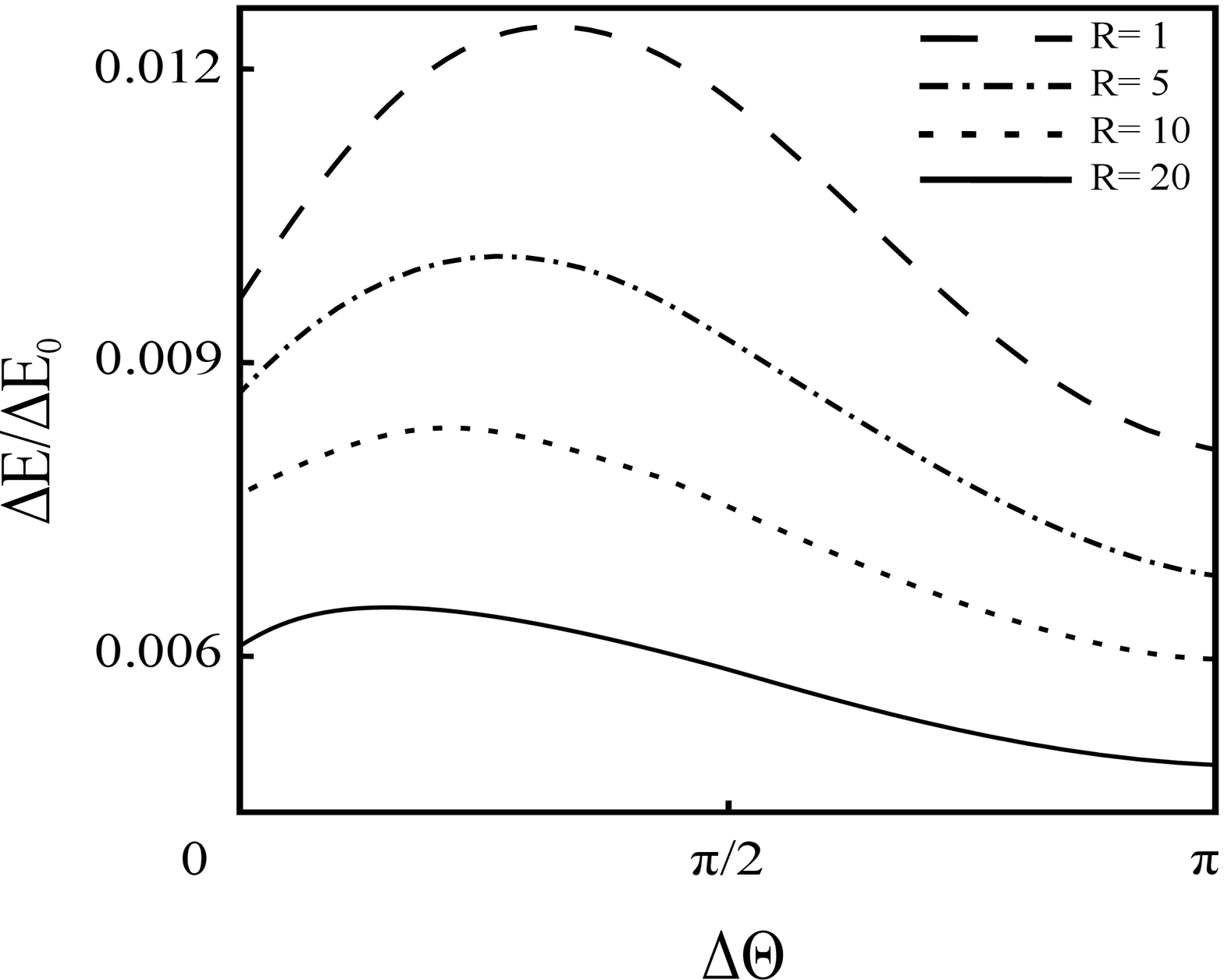}
		\caption{In the limit of large radii, the interaction between two rods becomes mostly repulsive, approaching the planar membrane case of pure repulsion.}
		\label{figS2_RodsAttracsRepulsions}
	\end{figure}

	For the interactions between two infinite rods, the Green's function becomes:
	\begin{equation}
	G(\Theta) = \frac{1}{32 \pi } \Big[ \cos (\Theta ) \left(4 \text{Li}_2\left(e^{-i \Theta }\right)+4 \text{Li}_2\left(e^{i \Theta }\right)-11\right)+
	12 (\Theta- \pi) \sin (\Theta )\Big],
	\end{equation}
	where $\mathrm{Li_n} (z) = \sum_{m=1}^{\infty}\frac{z^m}{m^n}\text{\space} \left( \text{with } z \in \mathbb{C}\right)$ is the polylogarithm function. Like the interaction between two rings, the inclusions' attraction and repulsion strength depends on the radius of the tube (Fig.~\ref{figS2_RodsAttracsRepulsions}).
	
	\section{Point-like inclusions}
	Evaluating the summation series in Eq.~\ref{Greensfunction}, we obtain the Green's function for point like inclusions:
	\begin{widetext}
		\begin{align}
		G (\theta,\zeta) & =\frac{e^{-\frac{\zeta }{\sqrt{2}}} \left(\sin \left(\frac{\zeta }{\sqrt{2}}\right)+\cos \left(\frac{\zeta }{\sqrt{2}}\right)\right)}{4 \sqrt{2} \pi }+ \frac{e^{-\frac{1}{2} \left(2+\sqrt{2}\right) \zeta -i \theta }}{512 \pi } \nonumber\\
		& -\left(\left(115+64 \sqrt{2}\right) e^{\sqrt{2} \zeta }-64 \sqrt{2}+115\right) \left(1+e^{2 i \theta }\right)+e^{\zeta +i \theta }\left(32 \sqrt{2} \left(e^{\sqrt{2} \zeta }-1\right) \text{Li}_2\left(e^{-\zeta -i \theta }\right)\right. \nonumber\\
		& + 32 \sqrt{2} \left(e^{\sqrt{2} \zeta }-1\right) \text{Li}_2\left(e^{i \theta -\zeta }\right)+32 \left(e^{\sqrt{2} \zeta }+1\right) \text{Li}_3\left(e^{i \theta -\zeta }\right)+32 \left(e^{\sqrt{2} \zeta }+1\right) \text{Li}_3\left(e^{-\zeta -i \theta }\right) \nonumber\\
		& + 32 \sqrt{2} \left(e^{\sqrt{2} \zeta }-1\right) \text{Li}_4\left(e^{-\zeta -i \theta }\right)+32 \sqrt{2} \left(e^{\sqrt{2} \zeta }-1\right) \text{Li}_4\left(e^{i \theta -\zeta }\right)+40 \left(e^{\sqrt{2} \zeta }+1\right) \text{Li}_5\left(e^{-\zeta -i \theta }\right) \nonumber\\
		& + \left.\left.40 \left(e^{\sqrt{2} \zeta }+1\right) \text{Li}_5\left(e^{i \theta -\zeta }\right)+43 \left(e^{\sqrt{2} \zeta }+1\right) \text{Li}_7\left(e^{i \theta -\zeta }\right)+43 \left(e^{\sqrt{2} \zeta }+1\right) \text{Li}_7\left(e^{-\zeta -i \theta }\right)\right)\right).
		\end{align}
	\end{widetext}
	
	\section{Effect of Casimir forces}
\begin{figure}[ht]
		\includegraphics[width=0.48\textwidth]{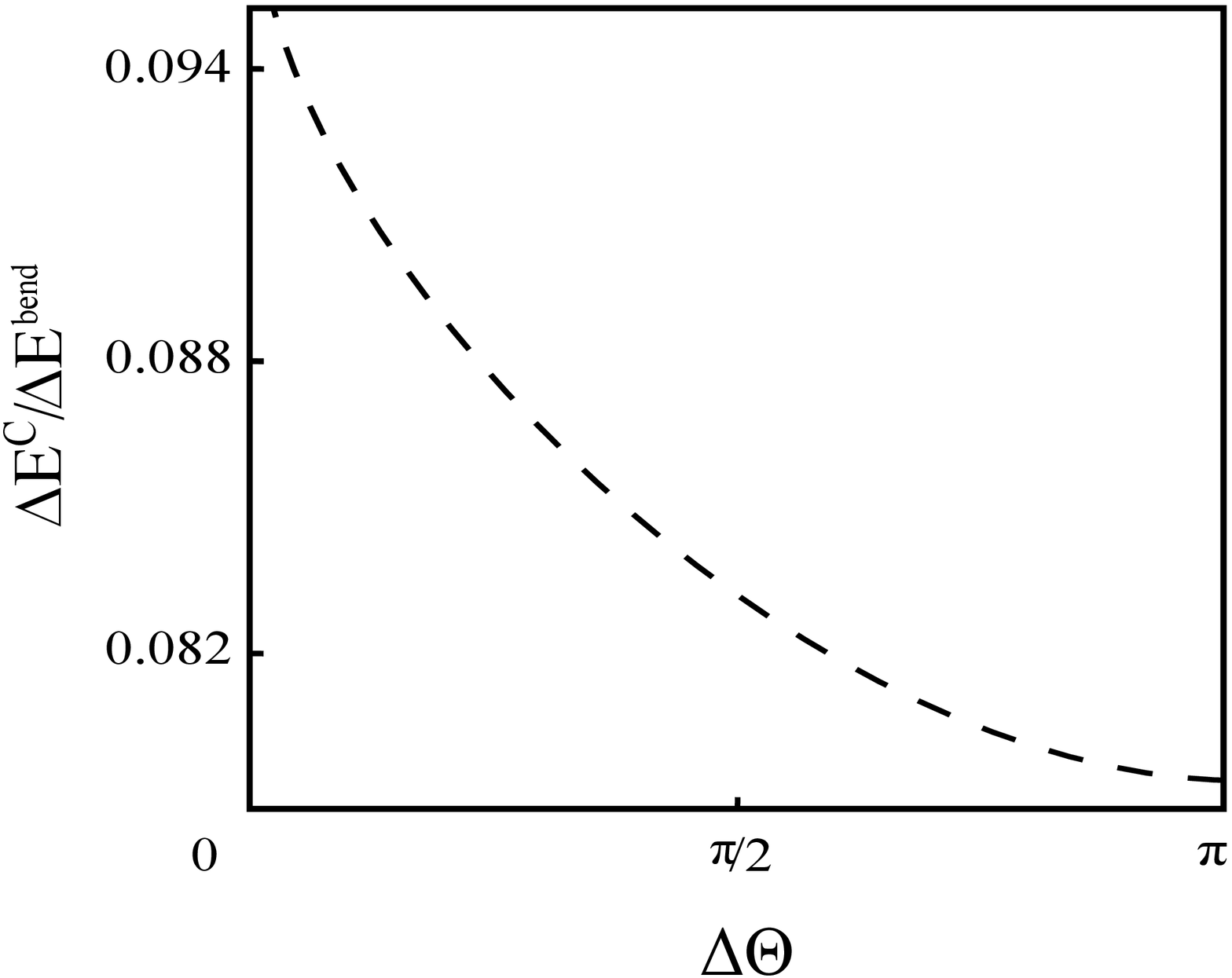}
		\caption{Ratio of fluctuation-induced energy to bending energy for an imposed curvature of $c = \frac{10}{R}$ and a bending modulus of $\kappa = 25 k_\mathrm{B} T$.}
		\label{figS3_Casimir}
	\end{figure}

	Membrane mediated interactions between inclusions like proteins embedded in a biological membrane originate from both the average deformation of the membrane and the constraints imposed on the fluctuations of the membrane. One can investigate the thermal fluctuation effects by constructing the canonical partition function of the fluctuation field (which in our work is parametrized by $u(\theta , \zeta )$) and applying the boundary conditions that are imposed by the inclusions. Following early work done by Ref. 43 of the main text, we would get exactly the same relation for the thermal energy:
	\begin{align}
	\Delta E^\mathrm{C} = \frac{k_B T}{2} \ln \det(M),
	\end{align}
	where $M$ is the matrix composed of derivatives of the Green's function that we derived for a membrane tube. Considering thermal Casimir effects, the total energy of the membrane becomes $\Delta E = \Delta E^{\mathrm{bend}} + \Delta E^\mathrm{C}$, where $\Delta E^{\mathrm{bend}}$ is the bending energy of the membrane, which is our primary interest in this work. As one can see, $\Delta E^\mathrm{C}$ depends only on the distance between the inclusions, which is hidden in the matrix $M$, and not the amount of curvature that is imposed. As illustrated in Fig.~\ref{figS3_Casimir}, the thermal effect is about an order of magnitude weaker than the mean-field contribution in the total energy of the membrane.
	
	\section{Monte Carlo simulations}
	During MC simulations, we simultaneously equilibrate two copies of our system in two different inverse temperatures ($\beta = \kappa/k_\mathrm{B} T$). For each copy of the system we use the Metropolis algorithm [41]: we accept any change in the configuration of our system with the probability $P[\Omega_{n} \rightarrow \Omega_{n+1}] = \min[1, \exp (-\beta \Delta E)]$. Global movements of inclusions are also allowed during simulations. The maximum step size of inclusions is adjusted such that acceptance rate of proposed moves is 50 $ \% $. In the end, in addition to (locally) minimizing the energy of the system in two different temperatures ($\beta_1 \text{ and } \beta_2$) separately, we also (globally) exchange the whole configurations corresponding to the temperatures based on the Metropolis algorithm: $P_\mathrm{Exch.} =\min[1, \exp \left (-(\beta_2-\beta_1) (E_2-E_1)\right )] $. More details about the method can be found in Refs. 41 and 42 of the main text.